# Effect of Interleaved FEC Code on Wavelet Based MC-CDMA System with Alamouti STBC in Different Modulation Schemes

Rifat Ara Shams[1], M. Hasnat Kabir[1], Sheikh Enayet Ullah[2]

[1]Department of Information and Communication Engineering, University of Rajshahi
Rajshahi-6205, Bangladesh.
swarna601@gmail.com
[2]Department of Applied Physics and Electronic Engineering, University of Rajshahi
Rajshahi-6205, Bangladesh.

## ABSTRACT

*In this paper, the impact of Forward Error Correction (FEC) code namely Trellis code with interleaver on the performance of wavelet based MC-CDMA wireless communication system with the implementation of Alamouti antenna diversity scheme has been investigated in terms of Bit Error Rate (BER) as a function of Signal-to-Noise Ratio (SNR) per bit. Simulation of the system under proposed study has been done in M-ary modulation schemes (MPSK, MQAM and DPSK) over AWGN and Rayleigh fading channel incorporating Walsh Hadamard code as orthogonal spreading code to discriminate the message signal for individual user. It is observed via computer simulation that the performance of the interleaved coded based proposed system outperforms than that of the uncoded system in all modulation schemes over Rayleigh fading channel.*

## KEYWORDS

## 1. INTRODUCTION

The most important objectives in the 4[th] generation wireless mobile communication systems are to enable the integration of existing technologies in a unified platform, to take care of the severe Inter-Symbol Interference (ISI), to provide high speed data transmission rate in a spectrally efficient manner with utilizing the available limited bandwidth and to achieve low cost and reduced complexity receivers and their signal processing solutions for high quality communication [1-2]. Multicarrier techniques on multi-path fading environment can be one of the best solutions to achieve the above performance. Due to the limitations of Orthogonal Frequency Division Multiplexing (OFDM) and Code Division Multiple Access (CDMA), Multi-Carrier Code Division Multiple Access (MC-CDMA) which is a multiple access scheme generated by the combination of OFDM and CDMA has received much attention [2]. Each user symbol in the frequency domain is spreaded by the MC-CDMA. In other word, each user symbol is phase shifted according to a code value and carried over multiple parallel sub-carriers. However, the code values differ per sub-carrier and per user. All sub-carrier signals are merged at the receiver and undo the code shift. Then the receiver separates signals of different users. So, the network can accommodate many users within a given frequency band [1,3]. It has many other advantages like robustness to channel dispersion, high frequency spectrum efficiency and high data transmission [2]. In Discrete Fourier Transform (DFT) based conventional MC-CDMA, the sub-carrier orthogonality and Synchronization are very sensitive to signal phase offset and frequency errors which will cause severe performance degradation due to the inter-channel interference (ICI), inter-symbol interference (ISI) and multiple access interference (MAI) [4,5]. For better performance, a number of improved MC-CDMA systems have been proposed. Among them, wavelet based MC-CDMA attracts much interests due to its powerful ability to combat multi-path interference (MPI) and ISI than conventional MC-CDMA [5]. Wavelet based MC-CDMA avoids the influence of delayed waves and eliminates the ISI by ensuring smaller sub channel bandwidth than the channel coherence bandwidth and by using a guard interval in multi-path environment [6]. In our previous work, the performance of the Wavelet based MC-CDMA system has been investigated using antenna diversity such as Alamouti STBC scheme [6] according to the suggestion of Salih M. [7]. Therefore, it is required to judge the performance of the system after implementing the Forward Error Correction (FEC) Code

In this paper, we propose a wavelet based MC-CDMA system using Forward Error Correction (FEC) with interleaving in different modulation schemes on fading environment. We consider error control coding technique to compensate the performance loss due to narrower bandwidth caused by multi-carrier technique. The reason for choosing FEC is to reduce the receiver complexity, eliminate the MAI effectively and give the receiver an ability to correct errors without needing a reverse channel to request retransmission of data [8]. Nevertheless, in this system, interleave can increase the performance by arranging data in a non-contiguous way. Interleave spreads the source bits out in time. Therefore, the important bits from a block of source data are not corrupted due to noise burst [9].

## 2. FORWARD ERROR CORRECTION USING TRELLIS CODE

In telecommunication, information theory and coding theory, forward error correction (FEC) is a technique used for controlling errors in data transmission caused by corruption from the unreliable or noisy communication channels, whereby the sender adds systematically generated redundant data to its messages, also known as an error-correcting code (ECC). The American mathematician Richard Hamming pioneered this field in the 1940s and invented the first error-correcting code, the Hamming code in 1950 [10].

Automatic repeat request (ARQ) and FEC mechanisms are well-known and widely used error detection and correction mechanisms for data transmission. In ARQ mechanism, if the sender fails to receive an acknowledgement before the timeout, it usually retransmits the lost packets to the receiver until the sender receives the explicit retransmission request from the receiver within certain expiration time or exceeds a predefined number of retransmissions. This mechanism is bad for multicast and is impracticable because of the retransmission delay [11]. In contrast to the ARQ mechanism, the FEC mechanism is considered more suitable for error correction. In a communication system that employs FEC, the information source sends a data sequence to an encoder and the encoder inserts redundant bits using a predetermined algorithm. This process generates a longer sequence of code bits, called a codeword, and such codeword is then transmitted to a receiver, which uses a suitable decoder to retrieve the original data sequence either during the process of transmission or on storage [12].

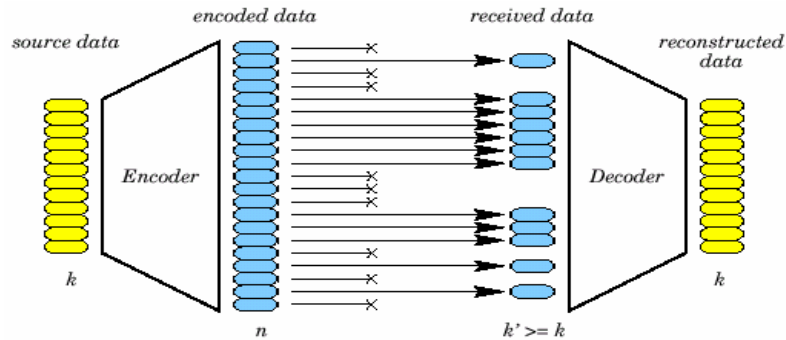

Figure 1: Block diagram of the operation of Forward Error Correction (FEC)

As shown in figure 1, the FEC encodes k packets with (n-k) redundant packets to form a FEC block with the n packets at the sender. Then the FEC can tolerate the loss of (n-k) packets in networks and recover the k packets with the help of the decoder from the FEC block at the receiver. It is advantageous that codes which introduce a large number of redundancy convey relatively little information per individual code bit, because it reduces the probability that all of the original data will be wiped out during transmission [10]. FEC has other advantages like it avoids multicast problems, sometimes no feedback channel is necessary, has long delay path and is one-way transmission. But a major disadvantage of it is the increase in channel bandwidth. Trellis coded modulation (TCM) is a combined binary convolution codes of rates (r, r+1) and modulation technique which introduces forward error correction coding without increasing the bandwidth of the channel signal. TCM, invented by Gottfried Ungerboeck in 1970s [13], is highly efficient technique because it provides high coding gain and little descending data rate by coding and error correction.

## 3. ALAMOUTI STBC SCHEME

Space time block coding (STBC) is a technique used in wireless communications to transmit multiple copies of a data stream simultaneously across different transmit antennas and to exploit the various received versions of the data to improve the transmission reliability, to achieve a full diversity order and to enjoy the simple maximum likelihood decoding [14-15]. For scattering environment the STBC is a practical, effective and widely applied technique for mitigating the effect of multipath fading because the transmitted signal must traverse a difficult environment with scattering, reflection, refraction and so on, and may then be further corrupted by thermal noise in the receiver means that some of the received copies of the data will be better than others, then STBC combines all the copies of the received signal in an optimal way to extract as much information from each of them as possible [14].

The Alamouti code, suggested by Mr. Siavash M Alamouti in his landmark October 1998 paper, is the only STBC that can achieve its full diversity gain using two transmit antennas and one receive antenna without needing to sacrifice its data rate and can provide the same diversity order as maximum-ratio receiver combining (MRRC) with two antennas at the receiver [16].

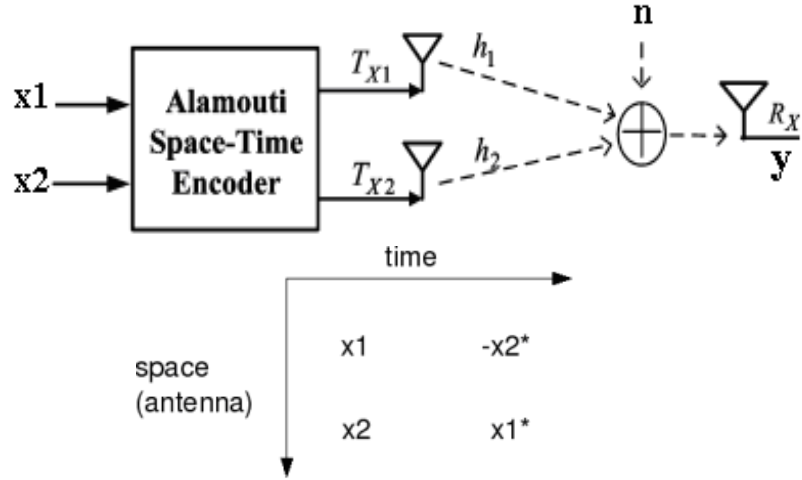

Figure 2: Conceptual block diagram of Alamouti STBC scheme

Figure 2 shows a wireless communication system with two transmit antennas and a single receive antenna. The transmitted symbols from the two transmitter antennas arrive at the receiver via two different channels and the discrete baseband received signal can be written as

$$y = h_1 x_1 + h_2 x_2 + n$$

where **y** is the received signal, $\mathbf{h_1}$ and $\mathbf{h_2}$ are the respective channel coefficients from the transmitter antenna 1 and 2 to the receive antenna, $\mathbf{x_1}$ and $\mathbf{x_2}$ are two corresponding transmitted symbols from these two antennas and **n** is a circularly symmetric complex Gaussian noise with variance $\sigma^2$ [17].

Let us consider, each set of two transmitted symbols spans over two consecutive time slots. In the first time slot, the received signal is,

$$y_1 = h_1 x_1 + h_2 x_2 - n_1 = \begin{bmatrix} h_1 & h_2 \end{bmatrix} \begin{bmatrix} x_1 \\ x_2 \end{bmatrix} + n_1$$

In the second time slot, the received signal is,

$$y_2 = -h_1 x_2^* + h_2 x_1^* - n_2 = \begin{bmatrix} h_1 & h_2 \end{bmatrix} \begin{bmatrix} -x_2^* \\ x_1^* \end{bmatrix} - n_2$$

Where $\mathbf{y_1}$, $\mathbf{y_2}$ is the received symbol on the first and second time slot respectively and $\mathbf{n_1}$, $\mathbf{n_2}$ is the noise on first and second time slots respectively.

For convenience, the above equation can be represented in matrix notation as follows:

$$\begin{bmatrix} y_1 \\ y_2^* \end{bmatrix} = \underbrace{\begin{bmatrix} h_1 & h_2 \\ h_2^* & -h_1^* \end{bmatrix}} \begin{bmatrix} x_1 \\ x_2 \end{bmatrix} + \begin{bmatrix} n_1 \\ n_2^* \end{bmatrix}$$

Let us define

$$\boldsymbol{H} = \begin{bmatrix} h_1 & h_2 \\ h_2^* & -h_1^* \end{bmatrix}$$

To solve for $\begin{bmatrix} x_1 \\ x_2 \end{bmatrix}$ , we know that we need to find the inverse of **H**.

For a general m x n matrix, the pseudo inverse is defined as,

$$H^+ = (H^H H)^{-1} H^H$$

The term,

$$(H^H H) = \begin{bmatrix} h_1^* & h_2 \\ h_2^* & -h_1 \end{bmatrix} \begin{bmatrix} h_1 & h_2 \\ h_2^* & -h_1^* \end{bmatrix} = \begin{bmatrix} |h_1|^2 - |h_2|^2 & 0 \\ 0 & |h_1|^2 + |h_2|^2 \end{bmatrix}$$

Since this is a diagonal matrix, the inverse is just the inverse of the diagonal elements, i.e.

$$(H^H H)^{-1} = \begin{bmatrix} \frac{1}{|h_1|^2 + |h_2|^2} & 0 \\ 0 & \frac{1}{|h_1|^2 + |h_2|^2} \end{bmatrix}$$

The estimate of the transmitted symbol is,

$$\begin{bmatrix} \widehat{x_1} \\ \widehat{x_2} \end{bmatrix} = (H^H H)^{-1} H^H \begin{bmatrix} y_1 \\ y_2^* \end{bmatrix}$$

$$= (H^H H)^{-1} H^H \left( H \begin{bmatrix} x_1 \\ x_2 \end{bmatrix} + \begin{bmatrix} n_1 \\ n_2^* \end{bmatrix} \right)$$

$$= \begin{bmatrix} x_1 \\ x_2 \end{bmatrix} + (H^H H)^{-1} H^H \begin{bmatrix} n_1 \\ n_2^* \end{bmatrix}$$

The bit error rate is,

$$P_{e,STBC} = p_{STBC}^2 \left[ 1 + 2(1 - p_{STBC}) \right]$$

Where,

$$p_{STBC} = \frac{1}{2} - \frac{1}{2} \left( 1 + \frac{2}{E_b / N_0} \right)^{-1/2}$$

Since ($H^H H$) is a diagonal matrix, hence there is no cross talk between x1, x2 after the equalizer [16].

## 4. SYSTEM MODEL

We consider, the wavelet based MC-CDMA transmitter includes M-branches and each of them consists of an up-sampler followed by a synthesis filter. The impulse response of this filter is derived from the wavelet orthogonal condition which generates a specific wavelet pulse. Here we propose conventional simulated wavelet based MC-CDMA system with interleaved forward error correction and implementation of Alamouti diversity scheme over AWGN and Rayleigh fading channel using different modulation techniques which is depicted in Figure 3.

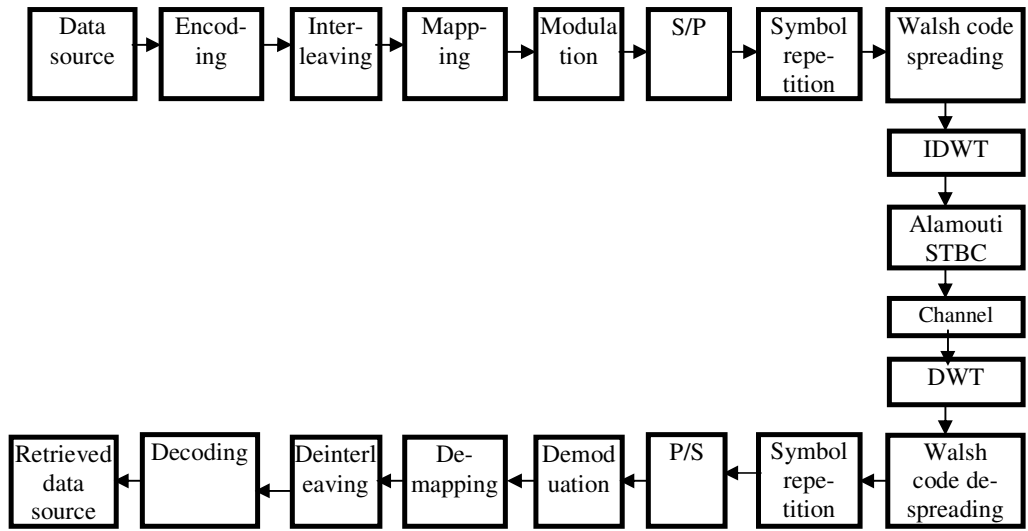

Figure 3: Block diagram of proposed wavelet based MC-CDMA system with implementation of FEC and Alamouti STBC scheme.

The proposed system with $Nc$ subcarriers and $K$ users, all with the same spreading factor $G$, is considered. Each user is assumed to have similar encoding, modulation and spreading. User data is frequency-division multiplexed over different orthogonal subcarriers, whereas a separate signature code sequence based on pseudo noise codes is assigned to each user to enable user separation. Here the synthetically generated binary bit stream for four users are fed into encoder to produce the required sequence of binary output vectors and then interleaver to obtain time diversity in a digital communications system without adding any overhead. Its output bits are mapped to a stream of coded symbols and the data are then converted into complex symbols using digital modulation and these symbols are converted into serial to parallel form and fed into copier section and multiplied with assigned individual Walsh Hadamard code. The coded symbols are fed into inverse discrete wavelet transformation whose output is a series of wavelet coefficients. Then the output is fed into the Alamouti space time block encoder whose output is then passed through the channel. Next the inverse operations are performed to retrieve the original data source signal at the receiver. Here the bit error rate (BER) performance as a function of signal-to-noise ratio (SNR) is examined.

$$SNR = 10\,log_{10}\left(\frac{\sigma_x{}^2}{\sigma_e{}^2}\right)$$

Where $\sigma_x{}^2$ is the mean square of the baseband signal and $\sigma_e{}^2$ is the mean square difference between the original and reconstructed signals.

## 5. SIMULATION RESULTS AND DISCUSSION

In our study, a simulation for MC-CDMA based on DWT has been made by using MATLAB 7.5. The developed program provides graphical display of various waveforms generated by the model. It calculates the bit error rate (BER) and also capable for simultaneous representation of both transmitted and received message signals. The simulation parameters are shown in Table 1. In different modulation schemes the bit error probability performance of DWT based MC-CDMA system is compared over AWGN channel. Moreover, the bit error probability performance of coded DWT based MC-CDMA system is determined and compared with the respective uncoded system under the fading environment namely Rayleigh fading channel in different modulation schemes.

Table 1: The parameters of the simulation model

| Parameter | Parameter Values |
|---|---|
| User | 4 |
| No. of bits used for synthetic data | 10,000 |
| SNR | 0-10 db |
| Mother wavelet | Daubechies |
| Modulation and Demodulation | BPSK, QPSK,4QAM, DBPSK |
| Spreading code | Walsh-Hadamard Code |
| FEC code | Trellis code |
| Wireless channel | AWGN channel, Rayleigh fading channel |
| Processing gain | 8 |

Figure 4 shows the comparison of the performance of different modulation schemes (BPSK, QPSK, 4QAM and DBPSK) for wavelet based MC-CDMA system in the presence of Additive White Gaussian Noise with implementation of Alamouti antenna diversity scheme for wide range of SNR from 0dB to 10dB. From the figure it is seen that the bit error rate (BER) is inversely proportional to the signal-to-noise ratio (SNR) as expected. At low SNR, BER reduction rate is small but with the increase of SNR the rate of BER reduction increases dramatically. In comparison with other modulation scheme, 4QAM shows much better BER performance than that of others in this system. At 2 dB SNR, the BER value of 4 QAM is approximately 5, 8 and 20 times lower than BPSK, DBPSK and QPSK, respectively.

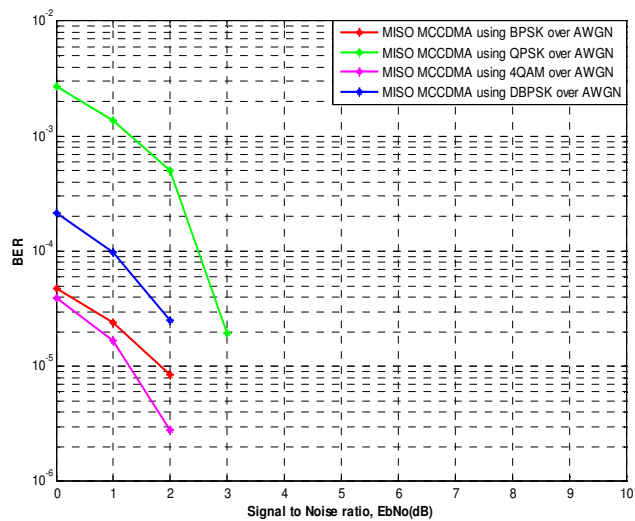

Figure 4: Comparison of different modulation schemes for Wavelet based MC-CDMA over AWGN channel.

To improve the performance of the proposed wireless communication system, to achieve the desired bit error probability and to minimize the burst error rate, interleaved FEC code can be used. The performance of interleaved trellis coded and uncoded wavelet based MC-CDMA system over Rayleigh fading channel using BPSK, QPSK, 4QAM and DBPSK modulation schemes is presented by figure 5, 6, 7 and 8 respectively.

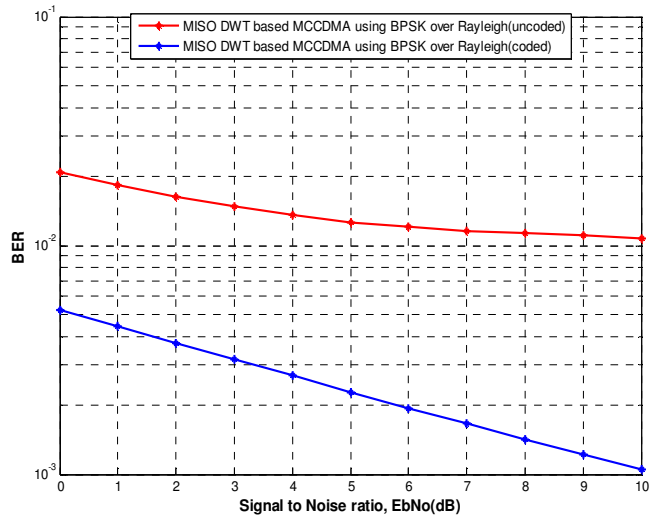

Figure 5: Comparison of coded and uncoded Wavelet based MC-CDMA using BPSK modulation scheme over Rayleigh fading channel.

From figure 5, it is seen that interleaved convolutional coding in discrete wavelet transform (DWT) based MC-CDMA with deployment of Alamouti diversity scheme using BPSK modulation can give better performance under Rayleigh fading channel. Contrast to uncoded DWT based system, coded one shows approximately 1 order of magnitude better BER performance at a typical SNR value of 10dB. Here, it is remarkable that the BER value of coded system decreases linearly with the increase of SNR.

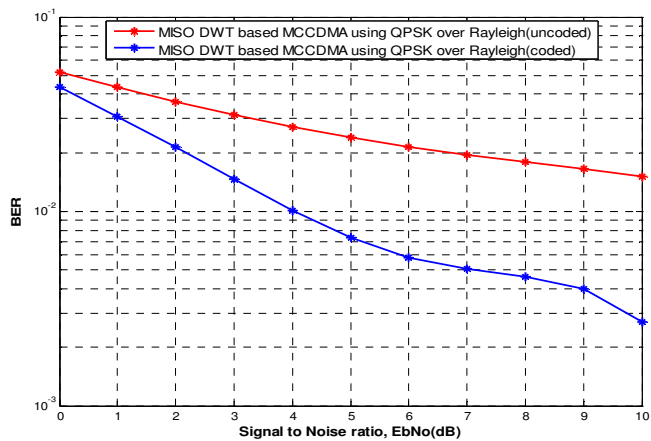

Figure 6: Comparison of coded and uncoded Wavelet based MC-CDMA using QPSK modulation scheme over Rayleigh fading channel.

The simulation result in figure 6 shows that interleaved coded DWT based MC-CDMA using QPSK modulation over Rayleigh fading channel has better BER performance than uncoded one as expected. For proposed coded communication system, when SNR is low, the rate of BER reduction is small but with the increase of SNR the BER reduction rate becomes high. At 3 dB SNR, the gain is roughly calculated which is almost 7 dB in the present system. On the other hand, it is not so significant al lower SNR.

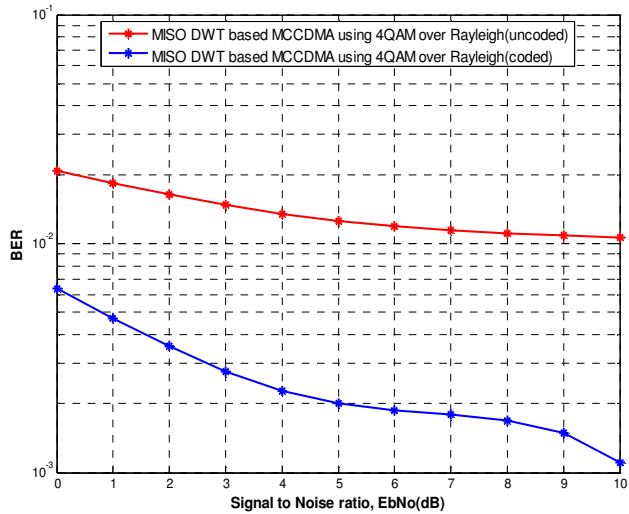

Figure 7: Comparison of coded and uncoded Wavelet based MC-CDMA using 4QAM modulation scheme over Rayleigh fading channel.

Figure 7 shows that, the coded proposed communication system using 4QAM modulation scheme gives better BER performance in contrast to uncoded system as expected. The rate of BER reduction for coded system is higher than that of uncoded one. The bit error probability 0.001 is obtained for coded proposed system at 10 dB SNR where as it is 0.01 for uncoded system at the same SNR. By comparing figure 4 and 7, it is found that the BER value of 4QAM modulation scheme is much lower in AWGN channel in contrast to Rayleigh fading channel.

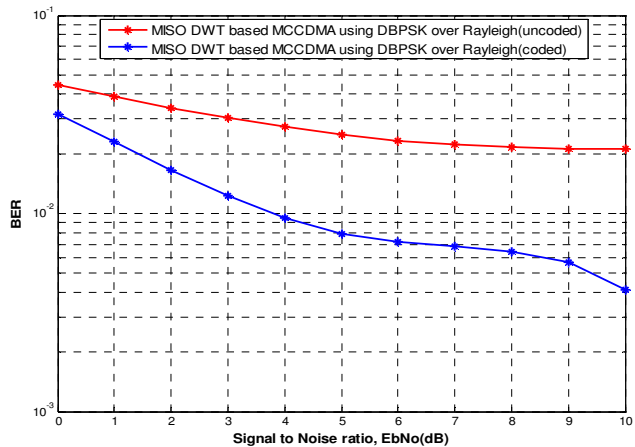

Figure 8: Comparison of coded and uncoded Wavelet based MC-CDMA using DBPSK modulation scheme over Rayleigh fading channel.

It is also seen from figure 8 that the BER performance of interleaved coded system using DBPSK modulation scheme is much better than that of uncoded one. A desirable gain which is slightly less than 9dB can be achieved from this proposed model using DBPSK.

## 6. CONCLUSION

In this paper the performance of wavelet based MC-CDMA system with Alamouti STBC using different modulation schemes over AWGN and Rayleigh fading channel has been studied. An approach has been presented that successfully applies FEC codes in the system to combat against the noise in channels and to improve the performance in contrast to uncoded system. It is seen that interleaved FEC coded wavelet based MC-CDMA system has better performance than that of uncoded system in fading channel irrespective of modulation schemes. From simulation results it is observed that in AWGN channel 4QAM modulation scheme has the best performance among others. On the other hand, in Rayleigh fading channel BPSK modulation technique provides the best performance among others though 4QAM has almost the same performance as BPSK in fading environment.